# Effect of Surrounding Conductive Object on Four-Plate Capacitive Power Transfer System

Qi Zhu, Lixiang Jackie Zou, Shaoge Zang, Mei Su, and Aiguo Patrick Hu

*Abstract-* In this paper, the effect of a surrounding conductive object on a typical capacitive power transfer (CPT) system with two pairs of parallel plates is studied by considering the mutual coupling between the conductive object and the plates. A mathematical model is established based on a 5*5 mutual capacitance matrix by using a larger additional conductive plate to represent the surrounding conductive object. Based on the proposed model, the effect of the additional conductive plate on the CPT system is analyzed in detail. The electric field distribution of the CPT system including the additional plate is simulated by ANSYS Maxwell. A practical CPT system consisting of four 100mm*100mm square aluminum plates and one 300mm*300mm square aluminum plate is built to verify the modeling and analysis. Both theoretical and experimental results show that the output voltage of the CPT system decreases when the additional conductive plate is placed closer to the CPT system. It has found that the additional plate can effectively shield the electric field outside the plate, and it attracts the electric field in-between the four plates of the CPT system and the additional plate. It has also found that the voltage potential difference between the additional plate and the reference plate of the CPT system remains almost constant even when the distance between them changes. The findings are useful for guiding the design of CPT systems, particularly the electric field shielding.

## I. INTRODUCTION

Capacitive Power Transfer (CPT) based on electric field coupling is an alternative Wireless Power Transfer (WPT) solution. It enables power transfer through metal barriers with lower power losses and EMI compared to Inductive Power Transfer (IPT) [1, 2]. This technology has been widely used in both low-power and high-power applications, such as integrated circuits [3], biomedical devices [4, 5], mobile devices [6, 7], synchronous machines [8-10] and electric vehicles [11-14]. Normally, a typical CPT system requires two pairs of parallel metal plates to form electric field coupling and provide a power flow path. Two metal plates are used in the primary side as power transmitters, and the other two are used in the secondary side as power receivers. The typical structure of a four-plate CPT system is shown in Fig. 1.

Based on the above typical structure, some fundamental studies have been done. C. Liu et al. studied the coupling mechanism [15], steady-state analysis [16], power flow control [17], 2D alignment analysis [18] and generalized coupling modeling [19] of a four-plate CPT system. L. Huang et al. studied compensation networks for improving the performance of a four-plate CPT system [20-21], the accurate steady-state modeling considering cross-coupling for a four-plate CPT system [22] and the definition of mutual coupling among four plates [23]. Four-plate CPT systems have been further developed for high-power applications. F. Lu et al. proposed a double-sided LCLC compensated four-plate CPT system to transfer 2.4kW power with 90.8% efficiency through an air gap distance of 150mm [11]. The typical structure of two pairs of parallel plates is replaced by a stacked four-plate structure to save space in electric vehicle charging applications, and the system achieved 1.88kW power with 85.87% efficiency through a 150mm air gap [12]. Two of the four plates are further replaced by the vehicle chassis and earth ground, with only two external plates required for electric vehicle charging [14]. Constant-voltage and constant-current modes were also found in a four-plate CPT system with double-sided LC compensation [24].

However in practice, a four-plate CPT system is quite sensitive to the surroundings, especially with a conductive object around it. The presence of this external object will change the existing electric field coupling generated by the four plates and further affect the system output performance. Theoretically, the original 4*4 mutual capacitance matrix contributed by the four plates should be correspondingly modified to a 5*5 matrix considering this external object. Hence, the effect of this external object on the four-plate system can be mathematically described by considering the additional elements in the 5*5 matrix. This study can be used to guide the practical design of CPT systems by considering the effects of surrounding conductive objects.

In this paper, the effect of a surrounding conductive object on a typical four-plate CPT system will be studied by considering the mutual coupling between this external object and one of the four plates. A larger additional conductive plate will be used to represent this external object, and a mathematical model based on a 5*5 mutual capacitance matrix will be established to describe the relationship between the additional plate and the CPT system. The effect of the additional plate on the CPT system will be analyzed in detail. The electric field distribution of the CPT system with an additional conductive plate will be simulated in ANSYS

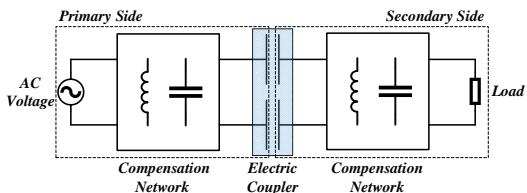

Fig. 1. Typical structure of a four-plate CPT system.

Maxwell. A practical CPT system with an additional conductive plate will be built to verify the modeling and analysis.

The rest of this paper is organized as follows: The modeling of a four-plate CPT system with a surrounding conductive object is introduced in Section II. The electric field distribution analysis with or without the surrounding conductive object is presented in Section III. Practical considerations related to the surrounding conductive object are given in Section IV. A four-plate CPT system with one surrounding conductive plate is established in Section V. A conclusion about the effect of a conductive object on a four-plate CPT system is drawn in Section VI.

## II. MODELING A FOUR-PLATE CAPACITIVE POWER TRANSFER SYSTEM WITH A SURROUNDING CONDUCTIVE OBJECT

A four-plate CPT system can be simplified as four metal plates with an AC voltage source and a load, regardless of the power converter and compensation network used in a typical CPT system. The placement of two pairs of parallel metal plates is taken to represent the capacitive coupler in a typical four-plate CPT system in this paper. Normally, the performance of a four-plate CPT system is sensitive to a nearby conductive object. Since some capacitance exists between any two electrical conductors in proximity, the presence of a nearby conductive object changes the electric field distribution of a four-plate CPT system. A larger metallic plate is used to represent the surrounding conductive object in this paper.

Assume that two pairs of metal plates are identical and placed in parallel in space, Plate 1, Plate 2, Plate 3 and Plate 4. A surrounding conductive object, Plate 5, is placed close to these four plates. The quantity of electric charge on each plate is $Q_1$, $Q_2$, $Q_3$, $Q_4$ and $Q_5$, respectively. The electric potential of each plate is $V_1$, $V_2$, $V_3$, $V_4$ and $V_5$, respectively. The mutual capacitance between every two plates is $C_{m,12}$, $C_{m,13}$, $C_{m,14}$, $C_{m,15}$, $C_{m,23}$, $C_{m,24}$, $C_{m,25}$, $C_{m,34}$, $C_{m,35}$ and $C_{m,45}$. $C_{m,ij}$ indicates how many charges on Plate $i$ are contributed by the potential difference between Plate $i$ and $j$. The self-capacitance of each plate is ignored because the distance between any two plates is much smaller than the distance between one plate and the ground reference at infinity. The block diagram of four plates and one surrounding conductive object is shown in Fig. 2.

The 5*5 mutual capacitance matrix of a four-plate CPT system with a surrounding conductive object can be described as follows:

$$\begin{bmatrix} Q_1 \\ Q_2 \\ Q_3 \\ Q_4 \\ Q_5 \end{bmatrix} = \begin{bmatrix} C_{m,1} & -C_{m,12} & -C_{m,13} & -C_{m,14} & -C_{m,15} \\ -C_{m,12} & C_{m,2} & -C_{m,23} & -C_{m,24} & -C_{m,25} \\ -C_{m,13} & -C_{m,23} & C_{m,3} & -C_{m,34} & -C_{m,35} \\ -C_{m,14} & -C_{m,24} & -C_{m,34} & C_{m,4} & -C_{m,45} \\ -C_{m,15} & -C_{m,25} & -C_{m,35} & -C_{m,45} & C_{m,5} \end{bmatrix} \begin{bmatrix} V_1 \\ V_2 \\ V_3 \\ V_4 \\ V_5 \end{bmatrix}$$
(1)

where,

$$C_{m,1} = C_{m,12} + C_{m,13} + C_{m,14} + C_{m,15}$$
$$C_{m,2} = C_{m,12} + C_{m,23} + C_{m,24} + C_{m,25}$$
$$C_{m,3} = C_{m,13} + C_{m,23} + C_{m,34} + C_{m,35}$$
$$C_{m,4} = C_{m,14} + C_{m,24} + C_{m,34} + C_{m,45}$$
$$C_{m,5} = C_{m,15} + C_{m,25} + C_{m,35} + C_{m,45}$$

Assume that Plate 1 and 2 are chosen as the primary power transmitters, and Plate 3 and 4 are chosen as the secondary power receiver. The source voltage, current and angular frequency are $V_S$, $I_S$ and $\omega$, respectively, and the voltage and current of the load $R_L$ are $V_L$ and $I_L$, respectively. The physical connection and equivalent circuit of a four-plate CPT system with a surrounding conductive object are shown in Fig. 3.

Let Plate 2 be the reference plate, then $V_2 = 0$, $V_s = V_1$, $V_L = V_3 - V_4$. According to the law of charge conservation, it can be drawn that $Q_1 + Q_2 + Q_3 + Q_4 + Q_5 = 0$. According to Kirchhoff's current law, (1) can be reduced to the following matrix:

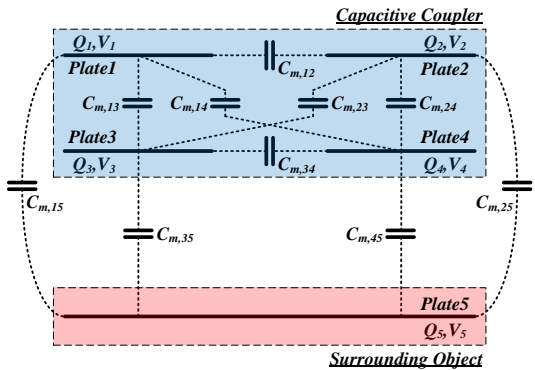

Fig. 2. Block diagram of four plates and one surrounding conductive object.

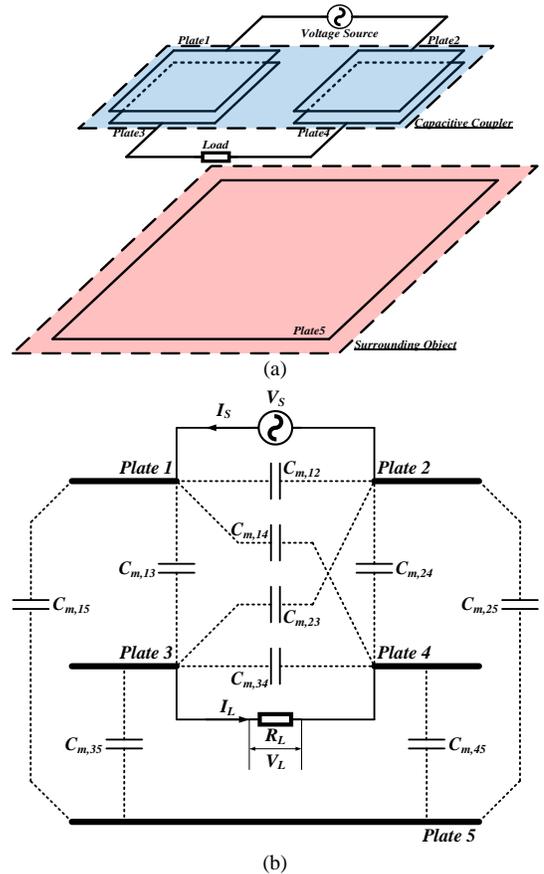

Fig. 3. (a) Physical connection and (b) equivalent circuit of a four-plate CPT system with a surrounding conductive object.

$$\begin{bmatrix} Q_1 \\ Q_2 \\ Q_3 \\ Q_4 \\ Q_5 \end{bmatrix} = \begin{bmatrix} I_S/j\omega \\ -I_S/j\omega \\ -I_L/j\omega \\ I_L/j\omega \\ 0 \end{bmatrix} = \begin{bmatrix} C_{m,1} & -C_{m,12} & -C_{m,13} & -C_{m,14} & -C_{m,15} \\ -C_{m,12} & C_{m,2} & -C_{m,23} & -C_{m,24} & -C_{m,25} \\ -C_{m,13} & -C_{m,23} & C_{m,3} & -C_{m,34} & -C_{m,35} \\ -C_{m,14} & -C_{m,24} & -C_{m,34} & C_{m,4} & -C_{m,45} \\ -C_{m,15} & -C_{m,25} & -C_{m,35} & -C_{m,45} & C_{m,5} \end{bmatrix} \begin{bmatrix} V_S \\ 0 \\ V_3 \\ V_4 \\ V_5 \end{bmatrix}$$

(2)

From the first, second and fifth rows in (2), the relationship between $V_S$, $I_S$, and $V_L$ can be expressed as (3). From the third, fourth and fifth rows in (2), the relationship between $V_L$, $I_L$, and $V_S$ can be expressed as (4). By moving $I_S$ and $I_L$ in (3) and (4) to the left-hand side, the relationship between the current and the voltage can be expressed as (5) and (6).

Normally, in a typical four-plate CPT system, the distance between Plate 1 and 3 (or Plate 2 and 4) is quite small compared to the distance between Plate 1 and 2, Plate 1 and 4, Plate 2 and 3, Plate 3 and 4. Meanwhile, the surrounding conductive object usually has a larger size compared to the size of the transmitter and receiver plates. For example, the surrounding object could be nearby electronic equipment or a human body. Hence, $C_{m,12}$, $C_{m,14}$, $C_{m,23}$ and $C_{m,34}$ are ignored in the following analysis.

(5) and (6) can be reduced as follows:
$$\begin{cases} I_S = j\omega C_1 V_S - j\omega C_M V_L \\ I_L = j\omega C_M V_S - j\omega C_2 V_L \end{cases} \quad (7)$$

where,

$$C_1 = \frac{C_{m,13}C_{m,24}(C_{m,15}+C_{m,25}+C_{m,35}+C_{m,45})}{(C_{m,13}+C_{m,24})(C_{m,15}+C_{m,25}+C_{m,35}+C_{m,45})+(C_{m,15}+C_{m,25})(C_{m,35}+C_{m,45})} + \frac{C_{m,13}C_{m,25}(C_{m,15}+C_{m,35}+C_{m,45})}{(C_{m,13}+C_{m,24})(C_{m,15}+C_{m,25}+C_{m,35}+C_{m,45})+(C_{m,15}+C_{m,25})(C_{m,35}+C_{m,45})} + \frac{C_{m,15}C_{m,24}(C_{m,25}+C_{m,35}+C_{m,45})}{(C_{m,13}+C_{m,24})(C_{m,15}+C_{m,25}+C_{m,35}+C_{m,45})+(C_{m,15}+C_{m,25})(C_{m,35}+C_{m,45})} + \frac{C_{m,15}C_{m,25}(C_{m,35}+C_{m,45})}{(C_{m,13}+C_{m,24})(C_{m,15}+C_{m,25}+C_{m,35}+C_{m,45})+(C_{m,15}+C_{m,25})(C_{m,35}+C_{m,45})}$$

$$C_2 = \frac{C_{m,13}C_{m,24}(C_{m,15}+C_{m,25}+C_{m,35}+C_{m,45})}{(C_{m,13}+C_{m,24})(C_{m,15}+C_{m,25}+C_{m,35}+C_{m,45})+(C_{m,15}+C_{m,25})(C_{m,35}+C_{m,45})} + \frac{C_{m,13}C_{m,45}(C_{m,15}+C_{m,25}+C_{m,35})}{(C_{m,13}+C_{m,24})(C_{m,15}+C_{m,25}+C_{m,35}+C_{m,45})+(C_{m,15}+C_{m,25})(C_{m,35}+C_{m,45})} + \frac{C_{m,24}C_{m,35}(C_{m,15}+C_{m,25}+C_{m,45})}{(C_{m,13}+C_{m,24})(C_{m,15}+C_{m,25}+C_{m,35}+C_{m,45})+(C_{m,15}+C_{m,25})(C_{m,35}+C_{m,45})} + \frac{C_{m,35}C_{m,45}(C_{m,15}+C_{m,25})}{(C_{m,13}+C_{m,24})(C_{m,15}+C_{m,25}+C_{m,35}+C_{m,45})+(C_{m,15}+C_{m,25})(C_{m,35}+C_{m,45})}$$

$$C_M = \frac{C_{m,13}C_{m,24}(C_{m,15}+C_{m,25}+C_{m,35}+C_{m,45})+C_{m,13}C_{m,25}C_{m,45}+C_{m,15}C_{m,24}C_{m,35}}{(C_{m,13}+C_{m,24})(C_{m,15}+C_{m,25}+C_{m,35}+C_{m,45})+(C_{m,15}+C_{m,25})(C_{m,35}+C_{m,45})}$$

Because $I_L = V_L/R_L$, the input impedance from the source side can be derived from the first equation of (7) when a conductive object is placed close to the CPT system:

$$Z_{in} = \frac{1}{j\omega C_1 + \dfrac{\omega^2 C_M^2}{\dfrac{1}{R_L}+j\omega C_2}}$$

(8)

The load voltage can be derived from the second equation of (7) when a conductive object is placed close to the CPT system:

$$V_L = \frac{j\omega C_M}{\dfrac{1}{R_L}+j\omega C_2} V_S$$

(9)

Hence, the load power can be expressed as follows when a conductive object is placed close to the CPT system:

$$P_L = \frac{\omega^2 C_M^2}{\left(\dfrac{1}{R_L}\right)^2 + \omega^2 C_2^2} \cdot \frac{|V_S|^2}{R_L}$$

(10)

Based on (8), (9) and (10), the system characteristics of a CPT system such as input impedance, output voltage and power will be affected by a surrounding conductive object. These equations can be used to quantitatively explain why the output of a four-plate CPT system fluctuates when a conductive object

$$V_S = \frac{I_S}{j\omega} \cdot \frac{C_{m,13}+C_{m,14}+C_{m,23}+C_{m,24}+\frac{(C_{m,15}+C_{m,25})(C_{m,35}+C_{m,45})}{C_{m,15}+C_{m,25}+C_{m,35}+C_{m,45}}}{\left(C_{m,13}+C_{m,14}+\frac{C_{m,15}(C_{m,35}+C_{m,45})}{C_{m,15}+C_{m,25}+C_{m,35}+C_{m,45}}\right)\left(C_{m,12}+\frac{C_{m,15}C_{m,25}}{C_{m,15}+C_{m,25}+C_{m,35}+C_{m,45}}\right)+(C_{m,23}+C_{m,24}+\frac{C_{m,25}(C_{m,35}+C_{m,45})}{C_{m,15}+C_{m,25}+C_{m,35}+C_{m,45}})\left(C_{m,12}+C_{m,13}+C_{m,14}+C_{m,15}-\frac{C_{m,15}C_{m,15}}{C_{m,15}+C_{m,25}+C_{m,35}+C_{m,45}}\right)} + V_L \\
\cdot \left[\frac{\left(C_{m,13}+\frac{C_{m,15}C_{m,35}}{C_{m,15}+C_{m,25}+C_{m,35}+C_{m,45}}\right)\left(C_{m,24}+\frac{C_{m,25}C_{m,45}}{C_{m,15}+C_{m,25}+C_{m,35}+C_{m,45}}\right)}{\left(C_{m,13}+C_{m,14}+\frac{C_{m,15}(C_{m,35}+C_{m,45})}{C_{m,15}+C_{m,25}+C_{m,35}+C_{m,45}}\right)\left(C_{m,12}+\frac{C_{m,15}C_{m,25}}{C_{m,15}+C_{m,25}+C_{m,35}+C_{m,45}}\right)+(C_{m,23}+C_{m,24}+\frac{C_{m,25}(C_{m,35}+C_{m,45})}{C_{m,15}+C_{m,25}+C_{m,35}+C_{m,45}})\left(C_{m,12}+C_{m,13}+C_{m,14}+C_{m,15}-\frac{C_{m,15}C_{m,15}}{C_{m,15}+C_{m,25}+C_{m,35}+C_{m,45}}\right)} \right.\\
\left. - \frac{\left(C_{m,14}+\frac{C_{m,15}C_{m,45}}{C_{m,15}+C_{m,25}+C_{m,35}+C_{m,45}}\right)\left(C_{m,23}+\frac{C_{m,25}C_{m,35}}{C_{m,15}+C_{m,25}+C_{m,35}+C_{m,45}}\right)}{\left(C_{m,13}+C_{m,14}+\frac{C_{m,15}(C_{m,35}+C_{m,45})}{C_{m,15}+C_{m,25}+C_{m,35}+C_{m,45}}\right)\left(C_{m,12}+\frac{C_{m,15}C_{m,25}}{C_{m,15}+C_{m,25}+C_{m,35}+C_{m,45}}\right)+(C_{m,23}+C_{m,24}+\frac{C_{m,25}(C_{m,35}+C_{m,45})}{C_{m,15}+C_{m,25}+C_{m,35}+C_{m,45}})\left(C_{m,12}+C_{m,13}+C_{m,14}+C_{m,15}-\frac{C_{m,15}C_{m,15}}{C_{m,15}+C_{m,25}+C_{m,35}+C_{m,45}}\right)}\right]$$

(3)

$$V_L = -\frac{I_L}{j\omega} \cdot \frac{C_{m,13}+C_{m,14}+C_{m,23}+C_{m,24}+C_{m,35}+C_{m,45}-\frac{(C_{m,35}+C_{m,45})(C_{m,35}+C_{m,45})}{C_{m,15}+C_{m,25}+C_{m,35}+C_{m,45}}}{\left(C_{m,13}+C_{m,23}+C_{m,34}+C_{m,35}-\frac{C_{m,35}C_{m,35}}{C_{m,15}+C_{m,25}+C_{m,35}+C_{m,45}}\right)\left(C_{m,14}+C_{m,24}+C_{m,34}+C_{m,45}-\frac{C_{m,45}C_{m,45}}{C_{m,15}+C_{m,25}+C_{m,35}+C_{m,45}}\right)-\left(C_{m,34}+\frac{C_{m,35}C_{m,45}}{C_{m,15}+C_{m,25}+C_{m,35}+C_{m,45}}\right)\left(C_{m,34}+\frac{C_{m,35}C_{m,45}}{C_{m,15}+C_{m,25}+C_{m,35}+C_{m,45}}\right)} + V_S \\
\cdot \frac{\left(C_{m,13}+\frac{C_{m,15}C_{m,35}}{C_{m,15}+C_{m,25}+C_{m,35}+C_{m,45}}\right)\left(C_{m,14}+C_{m,24}+C_{m,45}-\frac{(C_{m,35}+C_{m,45})C_{m,45}}{C_{m,15}+C_{m,25}+C_{m,35}+C_{m,45}}\right)-\left(C_{m,14}+\frac{C_{m,15}C_{m,45}}{C_{m,15}+C_{m,25}+C_{m,35}+C_{m,45}}\right)\left(C_{m,13}+C_{m,23}+C_{m,35}-\frac{C_{m,35}(C_{m,35}+C_{m,45})}{C_{m,15}+C_{m,25}+C_{m,35}+C_{m,45}}\right)}{\left(C_{m,13}+C_{m,23}+C_{m,34}+C_{m,35}-\frac{C_{m,35}C_{m,35}}{C_{m,15}+C_{m,25}+C_{m,35}+C_{m,45}}\right)\left(C_{m,14}+C_{m,24}+C_{m,34}+C_{m,45}-\frac{C_{m,45}C_{m,45}}{C_{m,15}+C_{m,25}+C_{m,35}+C_{m,45}}\right)-\left(C_{m,34}+\frac{C_{m,35}C_{m,45}}{C_{m,15}+C_{m,25}+C_{m,35}+C_{m,45}}\right)\left(C_{m,34}+\frac{C_{m,35}C_{m,45}}{C_{m,15}+C_{m,25}+C_{m,35}+C_{m,45}}\right)}$$

(4)

$$I_S = j\omega V_S \cdot \frac{\left(C_{m,13}+C_{m,14}+\frac{C_{m,15}(C_{m,35}+C_{m,45})}{C_{m,15}+C_{m,25}+C_{m,35}+C_{m,45}}\right)\left(C_{m,12}+\frac{C_{m,15}C_{m,25}}{C_{m,15}+C_{m,25}+C_{m,35}+C_{m,45}}\right)+(C_{m,23}+C_{m,24}+\frac{C_{m,25}(C_{m,35}+C_{m,45})}{C_{m,15}+C_{m,25}+C_{m,35}+C_{m,45}})\left(C_{m,12}+C_{m,13}+C_{m,14}+C_{m,15}-\frac{C_{m,15}C_{m,15}}{C_{m,15}+C_{m,25}+C_{m,35}+C_{m,45}}\right)}{C_{m,13}+C_{m,14}+C_{m,23}+C_{m,24}+\frac{(C_{m,15}+C_{m,25})(C_{m,35}+C_{m,45})}{C_{m,15}+C_{m,25}+C_{m,35}+C_{m,45}}} - j\omega V_L \\
\cdot \frac{\left(C_{m,13}+\frac{C_{m,15}C_{m,35}}{C_{m,15}+C_{m,25}+C_{m,35}+C_{m,45}}\right)\left(C_{m,24}+\frac{C_{m,25}C_{m,45}}{C_{m,15}+C_{m,25}+C_{m,35}+C_{m,45}}\right)-\left(C_{m,14}+\frac{C_{m,15}C_{m,45}}{C_{m,15}+C_{m,25}+C_{m,35}+C_{m,45}}\right)\left(C_{m,23}+\frac{C_{m,25}C_{m,35}}{C_{m,15}+C_{m,25}+C_{m,35}+C_{m,45}}\right)}{C_{m,13}+C_{m,14}+C_{m,23}+C_{m,24}+\frac{(C_{m,15}+C_{m,25})(C_{m,35}+C_{m,45})}{C_{m,15}+C_{m,25}+C_{m,35}+C_{m,45}}}$$

(5)

$$I_L = j\omega V_S \cdot \frac{\left(C_{m,13}+\frac{C_{m,15}C_{m,35}}{C_{m,15}+C_{m,25}+C_{m,35}+C_{m,45}}\right)\left(C_{m,14}+C_{m,24}+C_{m,45}-\frac{(C_{m,35}+C_{m,45})C_{m,45}}{C_{m,15}+C_{m,25}+C_{m,35}+C_{m,45}}\right)-\left(C_{m,14}+\frac{C_{m,15}C_{m,45}}{C_{m,15}+C_{m,25}+C_{m,35}+C_{m,45}}\right)\left(C_{m,13}+C_{m,23}+C_{m,35}-\frac{C_{m,35}(C_{m,35}+C_{m,45})}{C_{m,15}+C_{m,25}+C_{m,35}+C_{m,45}}\right)}{C_{m,13}+C_{m,14}+C_{m,23}+C_{m,24}+C_{m,35}+C_{m,45}-\frac{(C_{m,35}+C_{m,45})(C_{m,35}+C_{m,45})}{C_{m,15}+C_{m,25}+C_{m,35}+C_{m,45}}} - j\omega V_L \\
\cdot \frac{\left(C_{m,13}+C_{m,23}+C_{m,34}+C_{m,35}-\frac{C_{m,35}C_{m,35}}{C_{m,15}+C_{m,25}+C_{m,35}+C_{m,45}}\right)\left(C_{m,14}+C_{m,24}+C_{m,34}+C_{m,45}-\frac{C_{m,45}C_{m,45}}{C_{m,15}+C_{m,25}+C_{m,35}+C_{m,45}}\right)-\left(C_{m,34}+\frac{C_{m,35}C_{m,45}}{C_{m,15}+C_{m,25}+C_{m,35}+C_{m,45}}\right)\left(C_{m,34}+\frac{C_{m,35}C_{m,45}}{C_{m,15}+C_{m,25}+C_{m,35}+C_{m,45}}\right)}{C_{m,13}+C_{m,14}+C_{m,23}+C_{m,24}+C_{m,35}+C_{m,45}-\frac{(C_{m,35}+C_{m,45})(C_{m,35}+C_{m,45})}{C_{m,15}+C_{m,25}+C_{m,35}+C_{m,45}}}$$

(6)

such as a human body or electronic device is placed close to the CPT system.

Meanwhile, when a conductive object is placed close to the CPT system, its electric potential difference to the reference plate can be expressed as follows:

$$V_5 - V_2 = \left(\frac{C_\beta}{C_\gamma} + \frac{\frac{C_M C_\alpha}{C_\gamma}}{1 + j\omega C_2 R_L}\right) V_S \quad (11)$$

where,

$$C_\alpha = C_{m,13} C_{m,45} - C_{m,24} C_{m,35}$$
$$C_\beta = (C_{m,13} C_{m,15} + C_{m,13} C_{m,35} + C_{m,15} C_{m,35})(C_{m,24} + C_{m,45})$$
$$C_\gamma = C_{m,13} C_{m,24}(C_{m,15} + C_{m,25} + C_{m,35} + C_{m,45})$$
$$+ C_{m,13} C_{m,45}(C_{m,15} + C_{m,25} + C_{m,35})$$
$$+ C_{m,24} C_{m,35}(C_{m,15} + C_{m,25} + C_{m,45})$$
$$+ C_{m,35} C_{m,45}(C_{m,15} + C_{m,25})$$

$$\lim_{R_L \to 0} V_5 - V_2 = \frac{C_M C_\alpha + C_\beta}{C_\gamma} V_S$$

$$\lim_{R_L \to \infty} V_5 - V_2 = \frac{C_\beta}{C_\gamma} V_S$$

Based on (11), the electric potential difference between the surrounding conductive object and the reference plate of the CPT system is related to its relative position to the CPT system when the source and load are fixed. This relationship can be useful to identify the area under safe electric potential.

### III. ELECTRIC FIELD DISTRIBUTION ANALYSIS WITH THE PRESENCE OF A SURROUNDING CONDUCTIVE OBJECT

The previous section establishes the model of a four-plate CPT system with a surrounding conductive object based on a 5*5 mutual capacitance matrix. This section will analyze the effect of a surrounding conductive object on the electric field distribution of a four-plate CPT system by using ANSYS Maxwell. A simulation model consisting of two pairs of parallel 100mm*100mm square aluminum plates, and a 300mm*300mm square aluminum plate is built as shown in Fig. 4. The vertical distance between Plate 1 and 3 (or Plate 2 and 4) is set to be 10mm. The horizontal distance between Plate 1 and 2 (or Plate 3 and 4) is set to be 100mm. Plate 5 is placed in parallel with the plane of Plate 3 and 4, and the distance between the center of Plate 5 and the geometric center of Plate 3 and 4 is changed from 10mm to 100mm. The source voltage between Plate 1 and 2 is set to be 50Vrms@1MHz.

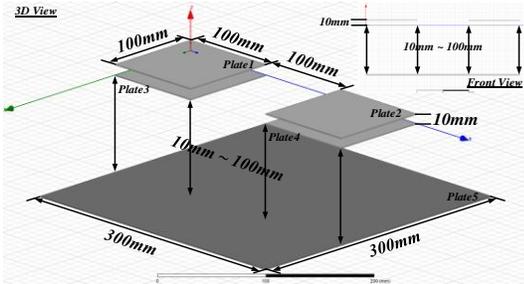

Fig. 4. Simulation model consisting of two pairs of parallel 100mm*100mm square aluminum plates, and a 300mm*300mm square aluminum plate.

Fig. 5 shows the electric field distribution of a four-plate CPT system without a surrounding conductive object at the instant $t=T/4$, where $T=1\mu s$.

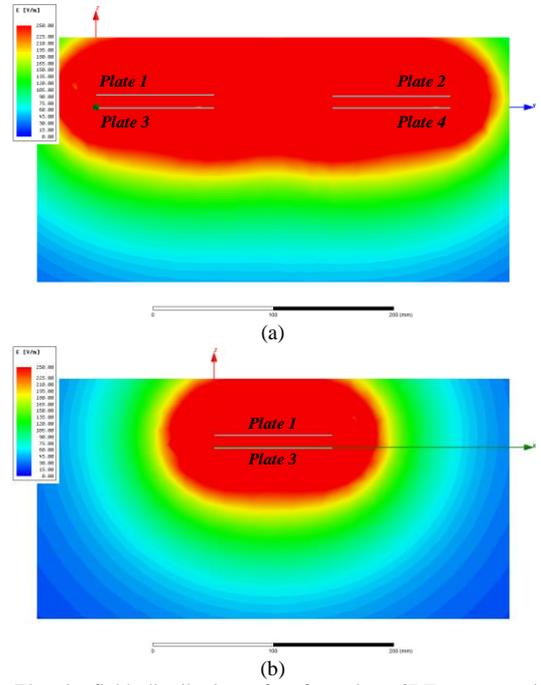

Fig. 5. Electric field distribution of a four-plate CPT system without a surrounding conductive object at the instant t=T/4, (a) Plane YOZ, (b) Plane XOZ.

Fig. 6 shows the electric field distribution of a four-plate CPT system when the plane of the surrounding conductive object is 100mm away from the plane of Plate 3 and 4 at the instant $t=T/4$, where $T=1\mu s$.

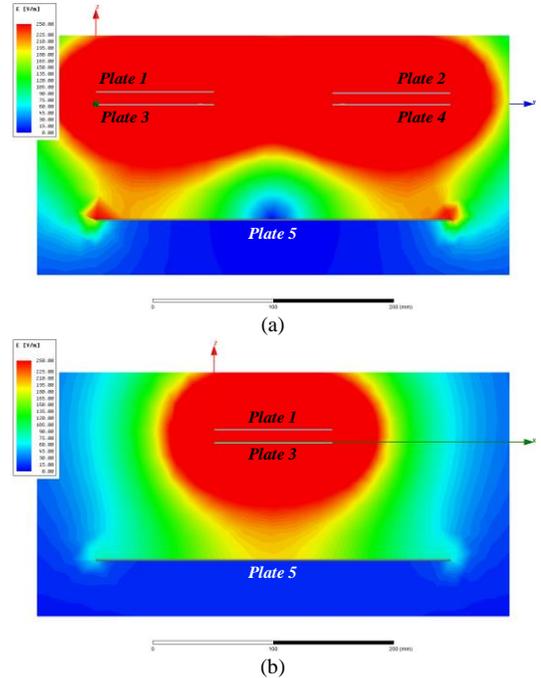

Fig. 6. Electric field distribution of a four-plate CPT system when Plate 5 is 100mm away from the plane of Plate 3 and 4 at the instant t=T/4, (a) Plane YOZ, (b) Plane XOZ.

Fig. 7 shows the electric field distribution of a four-plate CPT system when the plane of the surrounding conductive object is 50mm away from the plane of Plate 3 and 4 at the instant $t=T/4$, where $T=1\mu s$.

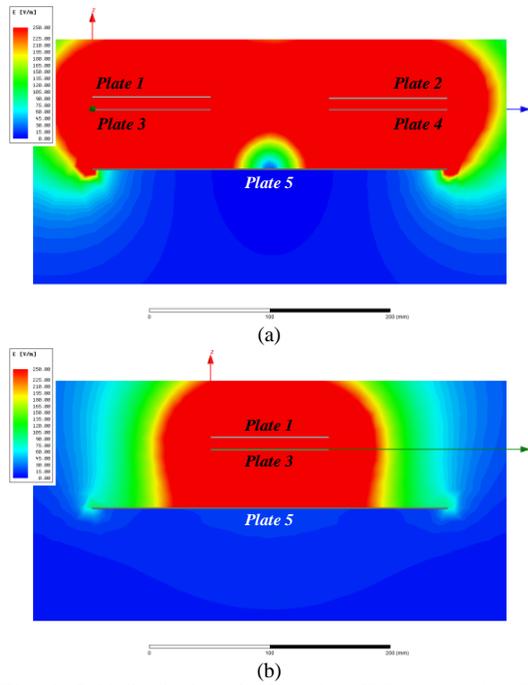

Fig. 7. Electric field distribution of a four-plate CPT system when Plate 5 is 50mm away from the plane of Plate 3 and 4 at the instant t=T/4, (a) Plane YOZ, (b) Plane XOZ.

Fig. 8 shows the electric field distribution of a four-plate CPT system when the plane of the surrounding conductive object is 10mm away from the plane of Plate 3 and 4 at the instant $t=T/4$, where $T=1\mu s$.

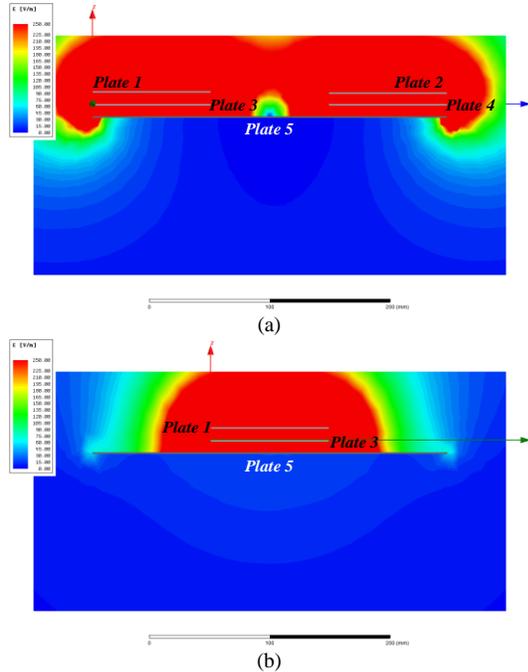

Fig. 8. Electric field distribution of a four-plate CPT system when Plate 5 is 10mm away from the plane of Plate 3 and 4 at the instant t=T/4, (a) Plane YOZ, (b) Plane XOZ.

TABLE I and Fig. 9 show the numerical values and the trend of the mutual capacitance between any two plates obtained from ANSYS Maxwell when the distance between the plane of Plate 5 and the plane of Plate 3 and 4 is changed from infinity to 10mm. It can be found from Fig. 9 that the curves of $C_{m,15}$ and $C_{m,25}$ are almost coincident, as well as the curves of $C_{m,35}$ and $C_{m,45}$. $C_{m,12}$, $C_{m,14}$, $C_{m,23}$ and $C_{m,34}$ are of small values and can be ignored.

TABLE I
SIMULATED MUTUAL CAPACITANCE BETWEEN EVERY TWO PLATES WITH THE DISTANCE BETWEEN THE CPT SYSTEM AND THE CONDUCTIVE OBJECT CHANGED

| Dist. (mm) | Mutual Capacitance (pF) | | | | | | | | | |
|---|---|---|---|---|---|---|---|---|---|---|
| | $C_{m,12}$ | $C_{m,13}$ | $C_{m,14}$ | $C_{m,15}$ | $C_{m,23}$ | $C_{m,24}$ | $C_{m,25}$ | $C_{m,34}$ | $C_{m,35}$ | $C_{m,45}$ |
| Inf. | 0.281 | 11.291 | 0.407 |  | 0.415 | 12.392 |  | 0.668 |  |  |
| 100 | 0.120 | 10.823 | 0.118 | 1.156 | 0.119 | 12.022 | 1.166 | 0.134 | 2.293 | 2.291 |
| 90 | 0.116 | 10.895 | 0.107 | 1.226 | 0.109 | 12.037 | 1.237 | 0.116 | 2.458 | 2.445 |
| 80 | 0.108 | 10.852 | 0.095 | 1.297 | 0.097 | 12.006 | 1.313 | 0.098 | 2.639 | 2.621 |
| 70 | 0.099 | 10.698 | 0.083 | 1.380 | 0.084 | 11.912 | 1.393 | 0.079 | 2.833 | 2.835 |
| 60 | 0.091 | 10.638 | 0.070 | 1.479 | 0.071 | 11.874 | 1.496 | 0.061 | 3.104 | 3.111 |
| 50 | 0.084 | 10.709 | 0.058 | 1.599 | 0.060 | 11.877 | 1.618 | 0.045 | 3.509 | 3.490 |
| 40 | 0.073 | 10.605 | 0.044 | 1.749 | 0.045 | 11.763 | 1.764 | 0.029 | 4.037 | 4.031 |
| 30 | 0.060 | 10.382 | 0.030 | 1.927 | 0.031 | 11.659 | 1.952 | 0.016 | 4.869 | 4.902 |
| 20 | 0.045 | 10.220 | 0.017 | 2.185 | 0.018 | 11.480 | 2.216 | 0.007 | 6.557 | 6.605 |
| 10 | 0.029 | 10.054 | 0.007 | 2.626 | 0.007 | 11.228 | 2.654 | 0.002 | 11.564 | 11.604 |

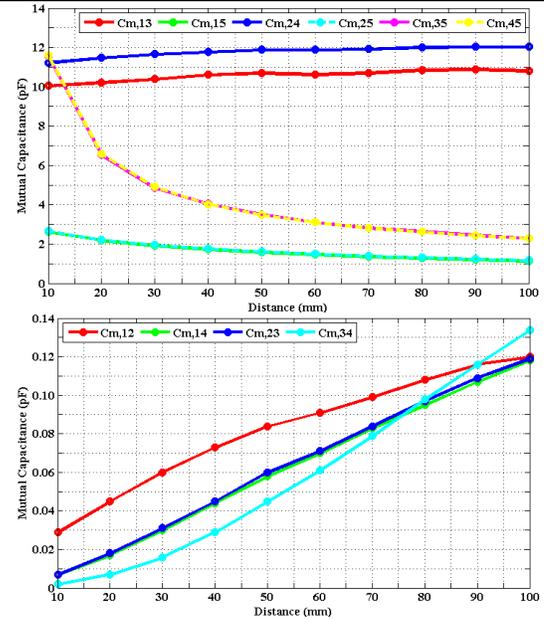

Fig. 9. Curves of the mutual capacitance between every two plates changes against the distance between the plane of Plate 5 and the plane of Plate 3 and 4.

Fig. 5 shows that the electric field distributions above and below the four plates are symmetric when no conductive object is near the four-plate CPT system. The range of the leakage electric field in Fig. 6 is larger than that in Fig. 5 when a conductive object is placed near the four-plate CPT system. Fig. 6(a) shows that the edges of the conductive object attract the electric field generated by the four-plate CPT system. With the decreasing distance between the conductive object and the four-plate CPT system, Fig. 7 shows that the conductive object starts to bound the leakage electric field. Fig. 8 shows that the leakage electric field below the four plates is shielded when the distance between the conductive object and the four-plate CPT system is almost equal to the distance between the transmitter and receiver plates. To summarize, the conductive object first attracts the electric field and then shields it when the distance between the conductive object and the CPT system is changed from 100mm to 10mm.

Fig. 9 shows that, when the conductive object is placed closer to the four-plate system, the coupling between the transmitter and receiver plates is almost constant. In contrast, the coupling between one of four plates and the conductive object increases. The resulting values of $C_{m,15}$, $C_{m,25}$, $C_{m,35}$ and $C_{m,45}$ will affect the system characteristics of the four-plate CPT system as discussed in the last section.

## IV. PRACTICAL CONSIDERATIONS RELATED TO THE SURROUNDING CONDUCTIVE OBJECT

To further simplify the above analysis for the following discussions, assume that $C_x=C_{m,13}=C_{m,24}$, $C_y=C_{m,15}=C_{m,25}$, $C_z=C_{m,35}=C_{m,45}$ according to TABLE I. Hence, $C_1$, $C_2$, $C_M$, $C_\alpha$, $C_\beta$, $C_\gamma$ can be rewritten as follows:

$$\begin{cases} C_1 = \dfrac{C_x C_x (C_y + C_z) + C_y C_y (C_x + C_z) + 2 C_x C_y C_z}{2 C_x C_y + 2 C_x C_z + 2 C_y C_z} \\ C_2 = \dfrac{C_x C_x (C_y + C_z) + (C_x + C_y) C_z C_z + 2 C_x C_y C_z}{2 C_x C_y + 2 C_x C_z + 2 C_y C_z} \\ C_M = \dfrac{C_x C_x (C_y + C_z) + C_x C_y C_z}{2 C_x C_y + 2 C_x C_z + 2 C_y C_z} \\ C_\alpha = C_x C_z - C_x C_z = 0 \\ C_\beta = (C_x C_y + C_x C_z + C_y C_z)(C_x + C_z) \\ C_\gamma = C_x C_x (C_y + C_y + C_z + C_z) + 2 C_x C_z (C_y + C_y + C_z) + C_z C_z (C_y + C_y) \end{cases}$$
(12)

By using (12), (9)-(11) can be reduced to the following equations:

$$\left|\frac{V_L}{V_S}\right| = \frac{\omega C_x}{\sqrt{\left(\frac{4}{R_L}\right)^2 + \omega^2 (C_x + C_z)^2}}$$
(13)

$$P_L = \frac{\omega^2 C_x^2}{\left(\frac{4}{R_L}\right)^2 + \omega^2 (C_x + C_z)^2} \cdot \frac{|V_S|^2}{R_L}$$
(14)

$$V_5 - V_2 = \frac{C_\beta}{C_\gamma} V_S = \frac{V_S}{2}$$
(15)

Based on (13) and (14), when the source and load are determined, the voltage gain and output power of the four-plate CPT system are affected by $C_x$, $C_z$. When the conductive object is placed closer, the value of $C_x$ is almost constant, therefore, $C_z$ will be the only factor affecting the system performance. It means the electric field coupling between the secondary receiver plate and the conductive object leads to changes in the voltage gain and output power. Based on (15), there exists an area that the electric potential difference between Plate 2 (the reference plate) and Plate 5 (the external conductive object) is equal to half of the source voltage. Within this area, $C_{m,12}$, $C_{m,14}$, $C_{m,23}$, $C_{m,34}$ are negligible compared to $C_{m,15}$, $C_{m,25}$, $C_{m,35}$, $C_{m,45}$. It also means that given a high enough voltage source, the electric potential difference of the conductive object in this area to the reference plate of the CPT system can be dangerous.

Normally, a conductive object can be mainly classified as either a deliberately added electric field shielding layer, or an unexpected foreign object with good conductivity. A piece of metal sheet, such as aluminum, copper, can be placed near a four-plate CPT system to shield the undesired electric field in one direction. Meanwhile, a working four-plate CPT system can be affected by an accidentally approaching conductive object, such as the human body, or electronic devices with metallic covers. The following discussions are based on the mutual effect between the conductive object and the CPT system.

### A. The effect of electric field shielding layer on the system performance

The deliberately added electric field shielding layer should affect the system performance as little as possible, while shielding the electric field. Besides the material of the shielding layer, its relative position to a four-plate CPT system can also affect the output performance of the CPT system according to (13) and (14). The following analysis focuses on the optimal $C_z$ with minimal impact on the system output performance when the source and load of the CPT system are determined.

The derivatives of (13) and (14) with respect to $C_z$ can be expressed as follows:

$$\frac{\partial}{\partial C_z} \left|\frac{V_L}{V_S}\right| = -\frac{\omega^3 C_x (C_x + C_z)}{\left[\left(\frac{4}{R_L}\right)^2 + \omega^2 (C_x + C_z)^2\right]^{\frac{3}{2}}} < 0$$
(16)

$$\frac{\partial P_L}{\partial C_z} = -\frac{2\omega^4 C_x^2 (C_x + C_z) R_L^3 |V_S|^2}{[16 + \omega^2 (C_x + C_z)^2 R_L^2]^2} < 0$$
(17)

According to (16) and (17), the voltage gain and output power decrease as $C_z$ increases, which means the electric field shielding layer will affect the output performance, and this effect will become larger as it is closer to the CPT system. According to TABLE I and (13), the output open-circuit voltage gain of the CPT system is half when there is a shielding layer 10mm away ($C_z \approx C_x$), compared to without any shielding layer ($C_z=0$). Hence, it is necessary to find out a suitable position for the shielding layer, where both the shielding requirement and system output performance are satisfied.

### B. The effect of electric field on the conductive foreign object

An unexpected foreign object with good conductivity not only affects the output performance of the CPT system, but also has a certain electric potential caused by the CPT system. This electric potential can be harmful to the foreign object itself, especially when the foreign object with good conductivity is the human body or an electronic device. TABLE II shows the potential difference between Plate 2 and 5 in Fig. 6-8 when Plate 5 is 10mm, 50mm and 100mm away from the plane of Plate 3 and 4. The simulated results are in good agreement with (15).

TABLE II
SIMULATED ELECTRIC POTENTIAL DIFFERENCE BETWEEN PLATE 2 AND 5 WHEN THE DISTANCE BETWEEN PLATE 5 AND THE PLANE OF PLATE 3 AND 4 IS DIFFERENT

| Distance (mm) | 10 | 50 | 100 |
|---|---|---|---|
| $V_1$-$V_2$ (Vrms@1MHz) | 50.00 | 50.00 | 50.00 |
| $V_5$-$V_2$ (Vrms@1MHz) | 24.59 | 24.89 | 24.99 |

The voltage applied between the two primary transmitter plates can be set sufficiently high to improve power transfer capacity of the CPT system. It means that the electric potential on the foreign object entering the defined area (where $C_{m,12}$, $C_{m,14}$, $C_{m,23}$, $C_{m,34}$ can be ignored compared to $C_{m,15}$, $C_{m,25}$, $C_{m,35}$, $C_{m,45}$) can be consequently high to the reference plate of the CPT system. Once the foreign object enters this area, regardless the distance between the foreign object and the CPT system, its electric potential difference to the reference plate of the CPT system is almost constant and can be dangerous. Hence, it is crucial in the four-plate system design stage to determine the safety area in which no object is allowed during operation.

## V. EXPERIMENTAL STUDY

A four-plate CPT system is shown in Fig. 10, which is composed of a signal generator Agilent 33220A for generating a high frequency voltage, a wideband power amplifier Agitek ATA-122D for amplifying the output voltage of the signal generator, an LCLC compensation network for further boosting the output voltage of power amplifier, four 100mm*100mm square aluminum plates as the transmitter and receiver plates, one 300mm*300mm square aluminum plate as the conductive object, and some different loads (1kΩ, 5kΩ and 10kΩ). In the experiments, the top two plates (Plate 1 and 2 as shown in Fig. 4) are connected to the output of the LCLC compensation network, and the middle two plates (Plate 3 and 4 as shown in Fig. 4) are connected to the resistor. The bottom plate (Plate 5 as shown in Fig. 4) is moved from infinity to 10mm away from the plane of Plate 3 and 4. The voltage applied between the top two plates is set to be 50Vrms@1MHz. It should be noticed that the signal generator and the power amplifier are powered by an isolation transformer to avoid connecting any plate directly to the ground, and a battery-powered handheld oscilloscope Keysight U1620A is used in the experiments to reduce the effect of the ground on the measurement results.

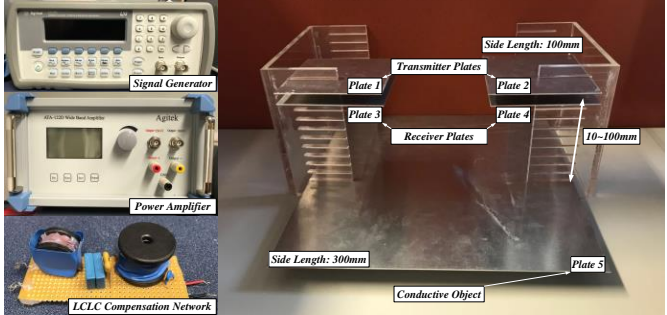

Fig. 10. A CPT system with four 100mm*100mm square aluminum plates as the transmitter and receiver plates, one 300mm*300mm square aluminum plate as the conductive object.

Ignoring $C_{m,12}$, $C_{m,14}$, $C_{m,23}$ and $C_{m,34}$, according to (9) and TABLE I, the calculated load voltages under different load conditions when the distance between the plane of Plate 5 and the plane of Plate 3 and 4 is changed from infinity to 10mm are shown in TABLE III.

TABLE III
CALCULATED LOAD VOLTAGES UNDER DIFFERENT LOAD CONDITIONS WHEN THE DISTANCE BETWEEN THE CPT SYSTEM AND THE CONDUCTIVE OBJECT VARYS

| Distance (mm) | Calculated Load Voltage (V) | | |
|---|---|---|---|
| | $R_L$=1kΩ | $R_L$=5kΩ | $R_L$=10kΩ |
| Inf. | 1.855 | 9.124 | 17.400 |
| 100 | 1.788 | 8.748 | 16.441 |
| 90 | 1.795 | 8.778 | 16.469 |
| 80 | 1.789 | 8.745 | 16.388 |
| 70 | 1.769 | 8.645 | 16.191 |
| 60 | 1.761 | 8.600 | 16.077 |
| 50 | 1.768 | 8.619 | 16.053 |
| 40 | 1.751 | 8.523 | 15.816 |
| 30 | 1.724 | 8.373 | 15.445 |
| 20 | 1.697 | 8.195 | 14.910 |
| 10 | 1.663 | 7.871 | 13.675 |

TABLE IV shows the measured load voltages under different load conditions when the distance between the plane of Plate 5 and the plane of Plate 3 and 4 is changed from infinity to 10mm.

TABLE IV
MEASURED LOAD VOLTAGES UNDER DIFFERENT LOAD CONDITIONS WHEN THE DISTANCE BETWEEN THE CPT SYSTEM AND THE CONDUCTIVE OBJECT VARYS

| Distance (mm) | Measured Load Voltage (V) | | |
|---|---|---|---|
| | $R_L$=1kΩ | $R_L$=5kΩ | $R_L$=10kΩ |
| Inf. | 3.101 | 10.961 | 15.585 |
| 100 | 3.072 | 10.902 | 15.340 |
| 90 | 3.030 | 10.829 | 15.220 |
| 80 | 2.917 | 10.770 | 15.143 |
| 70 | 2.892 | 10.669 | 15.098 |
| 60 | 2.834 | 10.511 | 15.035 |
| 50 | 2.805 | 10.462 | 14.904 |
| 40 | 2.780 | 10.310 | 14.640 |
| 30 | 2.768 | 10.138 | 14.429 |
| 20 | 2.752 | 9.839 | 13.871 |
| 10 | 2.686 | 9.170 | 12.631 |

Fig. 11 shows the curves of the calculated and measured load voltage against distance changing under different load conditions by using the data from TABLE III and IV, where the solid lines present the calculated values, and the scatter dots present the measured values. Red, green and blue present the values at 1kΩ, 5kΩ and 10kΩ.

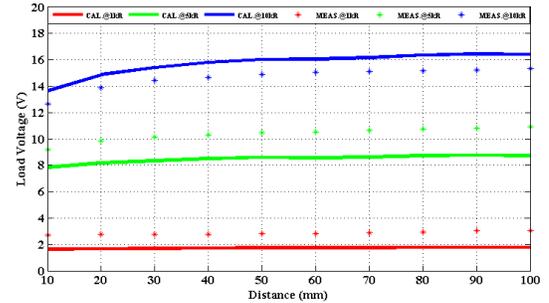

Fig. 11. Curves of the calculated and measured load voltage against distance change under different load conditions, where CAL. and MEAS. mean calculated and measured values.

Considering the differences between the simulation model as shown in Fig. 4 and the practical setup as shown in Fig. 10, the measured load voltages are in good agreement with the calculated load voltages. Both the calculated and measured load voltages show that the output performance of the four-plate CPT system degrades when the distance between the CPT system and the conductive object decreases. The error between the calculated and measured values can be caused by many factors, such as the slight misalignment of the four plates, the dimension bias of the four plates, the cables connected to the plates, the probes connected to the plates, the loads with poor high-frequency characteristic and the surroundings. All the above factors can change the mutual capacitance between any two plates.

Fig. 12 and 13 show the effect of two upper and lower shielding layers on the output performance of the four-plate CPT system. The voltage applied between the top two plates of the CPT system is increased to 100Vrms@1MHz, and the load is set to be 1kΩ.

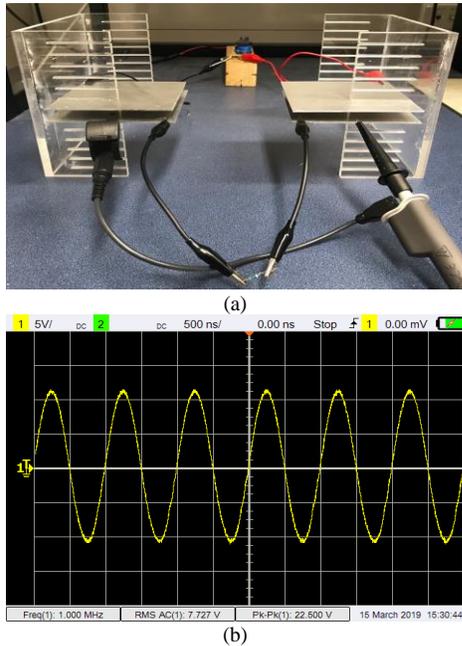

Fig. 12. (a) Practical setup and (b) measured load voltage without the upper and lower shielding layers.

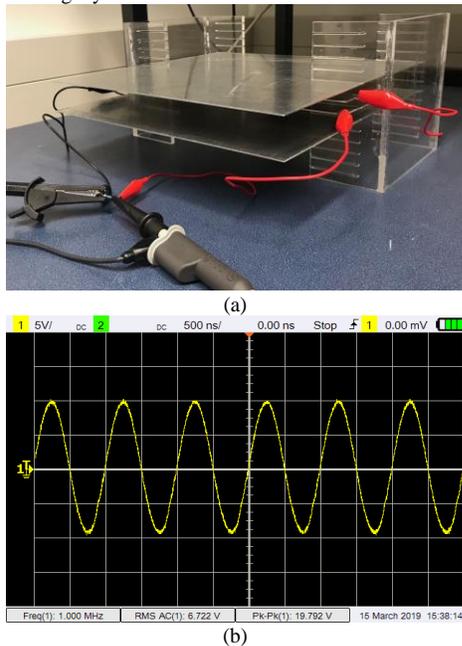

Fig. 13. (a) Practical setup and (b) measured load voltage with the upper and lower shielding layers.

From Fig. 12 and 13, it can be found that the load voltage will drop from 7.727V to 6.722V when the upper and lower shielding layers are added, which means the shielding layers will shield the electric field and degrade the system output performance at the same time.

Fig. 14 and 15 show the effect of a nearby human hand on the output performance of the four-plate CPT system. The voltage applied between the top two plates of the CPT system is increased to 100Vrms@1MHz, and the load is set to be 1kΩ.

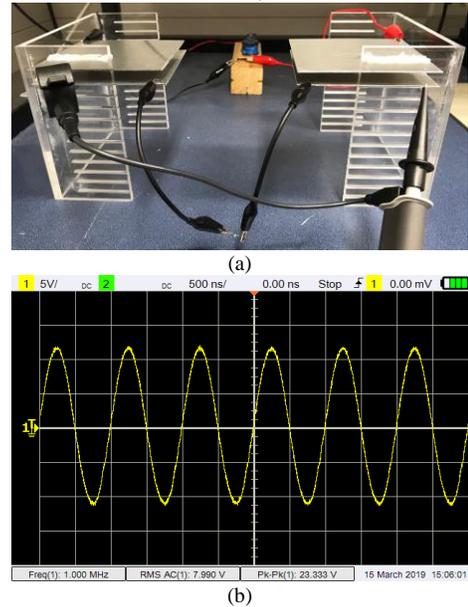

Fig. 14. (a) Practical setup and (b) measured load voltage without the nearby human hand.

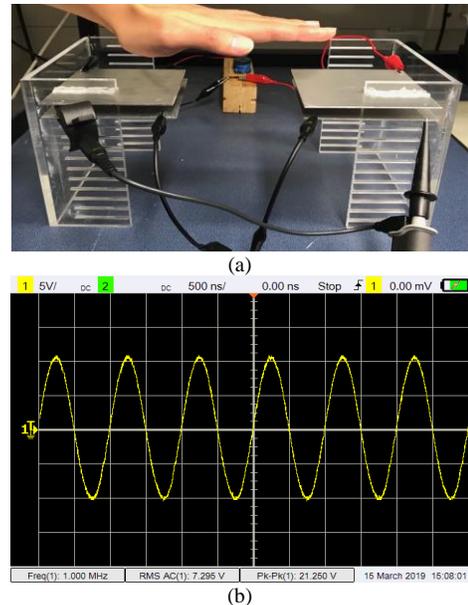

Fig. 15. (a) Practical setup and (b) measured load voltage with the nearby human hand.

From Fig. 14 and 15, it can be found that the load voltage will drop from 7.990V to 7.295V when the human hand is 30mm away from the plane of the top two plates, which means the unexpected object with good conductivity will affect the system output performance at the same time when it enters the charging area.

## VI. CONCLUSION

In this paper, the effect of a surrounding conductive object on a typical four-plate CPT system has been studied by considering the mutual coupling between the conductive object and the four plates into a 5*5 mutual capacitance matrix. A larger additional conductive plate has been used to represent the conductive object, and a mathematical model has been established to analyze the interaction between the conductive plate and the CPT system. The electric field distribution of the CPT system with the conductive plate has been simulated in ANSYS Maxwell. A practical system consisting of four 100mm*100mm square aluminum plates and one 300mm*300mm square aluminum plate has been built to verify the modeling and analysis. Both theoretical and experimental results have shown that the output voltage of the CPT system decreases when the conductive plate is placed closer to the CPT system. It has found that the conductive plate first attracts the electric field of the CPT system and then shields it when the distance between the conductive plate and the CPT system is changed from 100mm to 10mm. Meanwhile, the electric potential difference between the conductive plate and the reference plate of the CPT system remains almost constant. The findings from this research can be used to guide the practical design of CPT systems by taking the effects of surrounding conductive objects into consideration.